\documentclass[%
 reprint,
superscriptaddress,
 amsmath,amssymb,
 aps,
]{revtex4-1}

\usepackage{graphicx}
\usepackage{dcolumn}
\usepackage{bm}

\begin{document}

\preprint{APS/123-QED}

\title{Wood Anomalies and Surface-Wave Excitation with a Time-Grating}

\author{Emanuele Galiffi}
\affiliation{The Blackett Laboratory, Imperial College London, SW7 2AZ, London, UK}


\author{Yao-Ting Wang}
\affiliation{Department of Mathematics, Imperial College London, SW7 2AZ, London, UK}

\author{Zhen Lim}
\affiliation{The Blackett Laboratory, Imperial College London, SW7 2AZ, London, UK}

\author{J. B. Pendry}
\affiliation{The Blackett Laboratory, Imperial College London, SW7 2AZ, London, UK}

\author{Andrea Alù}
\affiliation{Photonics Initiative, Advanced Science Research Center,
City University of New York, New York, New York 10031, USA}

\author{Paloma A. Huidobro}
\affiliation{Instituto de Telecomunica\c c\~oes, Instituto Superior Tecnico-University of Lisbon, Portugal}




\date{\today}

\begin{abstract}
    In order to confine waves beyond the diffraction limit, advances in fabrication techniques have enabled subwavelength structuring of matter, achieving near-field control of light and other types of waves. The price is often expensive fabrication needs and the irreversibility of device functionality, as well as the introduction of impurities, a major contributor to losses. In this Letter, we propose temporal inhomogeneities, such as a periodic drive in the electromagnetic properties of a surface which supports guided modes, to circumvent the need for subwavelength fabrication in the coupling of propagating waves to evanescent modes across the light line, achieving the temporal counterpart of the Wood anomaly. We show analytically and numerically how this concept is valid for any material platform and at any frequency, and propose and model a realistic experiment in graphene to couple terahertz radiation to plasmons with unit efficiency, demonstrating that time-modulation of material properties could be a tunable, lower-loss and fast-switchable alternative to the subwavelength structuring of matter for near-field wave control.
\end{abstract}

\maketitle

The idea of trapping light near a surface finds its roots in the curious discovery by Wood that the reflection of light from a metallic grating vanishes at a specific angle of incidence, as diffraction theory alone seemed unable to explain \cite{wood1902remarkable}. Simultaneously, theories were developed, which predicted the existence of evanescent modes near metallic surfaces \cite{sommerfeld1899ueber, zenneck1907fortpflanzung}, although only half a century later the link between these two phenomena was elucidated \cite{fano1941theory}. This scattering anomaly was thus recognised as the coupling of the incoming photons to surface waves guided along a metal-dielectric interface by the free electron plasma \cite{ritchie1957plasma,hessel1965new,raether1988surface}.

Over the past decades, advances in nano-fabrication have enabled unprecedented progress in the harvesting and control of electromagnetic waves near surfaces or layered materials, with several types of surface excitations available, such as surface plasmon, phonon and exciton polaritons \cite{basov2016polaritons,low2017polaritons}. Similarly, in the RF and microwave regimes, the engineering of metasurfaces enables the realisation of effective impedance surfaces which support guided waves with tailored properties \cite{sievenpiper1999high,pendry2004mimicking}.

By their definition, surface waves are evanescent along at least one spatial dimension. The consequent imaginary out-of-plane momentum component, $k_z = \sqrt{\omega^2/c^2 - k_x^2}$, forces the in-plane momentum $k_x$ to be larger than the total momentum $\omega/c$ of any photon propagating in free space. Here $\omega$ and $c$ are the frequency of the wave and the speed of light in the medium. This implies that the coupling of these two waves is conditional to the bridging of this momentum mismatch, which needs to suffice to couple a propagating wave, within the light cone, to an evanescent one, outside of it. In free space this is conventionally done by breaking the translational invariance of the surface either by means of a near-field probe, such as a tip, or, as Wood did, by periodically structuring the surface of the metal, or its surroundings \cite{maier2007plasmonics}. Alternative schemes have been proposed, although they rely on inherently weak nonlinear processes \cite{renger2009free,constant2016all}.

While spatial design and fabrication has paved the way to the discovery and harnessing of novel near-field phenomena, the limited precision of spatial fabrication methods stands as a major obstacle to the decade-long challenge to reduce losses in nanophotonics. On the other hand, the recent advent of tunable and ultra-thin van der Waals layered materials, such as graphene, and thin-film semiconductors such as indium tin oxide \cite{basov2016polaritons,low2017polaritons,alam2016large}, has sparked the development of new schemes to tune and modulate the electromagnetic properties of a material in time, empowering the field of metamaterials and metasurfaces with the potential to exploit time as a new degree of freedom for the control of light \cite{shaltout2019spatiotemporal,bruno2020negative,datta2020low}. Similarly, at RF and microwave frequencies, active capacitive elements can be incorporated in a metasurface to achieve temporal and spatiotemporal control of electromagnetic waves \cite{sounas2017non,ramaccia2019phase,caloz2019spacetime,bruno2020negative,cardin2020surface,pachecopena2020temporal}, while similar efforts have been made in the acoustics \cite{fleury2016floquet}, elasticity \cite{darabi2020reconfigurable,trainiti2019time} and water-wave \cite{bacot2016time} communities, each featuring challenges and advantages specific to the physics involved.

In this Letter, we propose the temporal counterpart of the Wood anomaly as a new mechanism to excite surface waves without need for in-plane structuring. By periodically modulating the surface impedance of a thin film, we demonstrate that surface waves can be excited with unit efficiency from the far-field, with no spatial inhomogeneity on their surface, and suggest and model a realistic experiment based on a time-modulated graphene sheet.

Our results provide an alternative path towards the coupling of radiation to surface waves, which circumvents the need for near-field probes,  subwavelength grating structures and/or advanced fabrication. This concept bears the potential to dramatically reduce losses, as well as costs, in current nanophotonics setups, while simultaneously enabling fast switchability and wider tunability for metasurfaces across the electromagnetic spectrum, as well as in other wave systems.

This Letter is structured as follows: we first generalise the original concept of the Wood anomaly to the temporal dimension, demonstrating analytically and verifying numerically its capabilities for the trapping of radiation into surface waves for the case of a time-modulated, spatially homogeneous, general conducting layer. We then propose a realistic implementation in graphene, and show that critical coupling can be achieved with the inclusion of a back-reflector, as a result of the interaction between a cavity mode and a graphene plasmon enabled by the temporal grating.

\begin{figure}[t]
\vspace{10pt}
    \centering
    \includegraphics[width = 0.9\columnwidth]{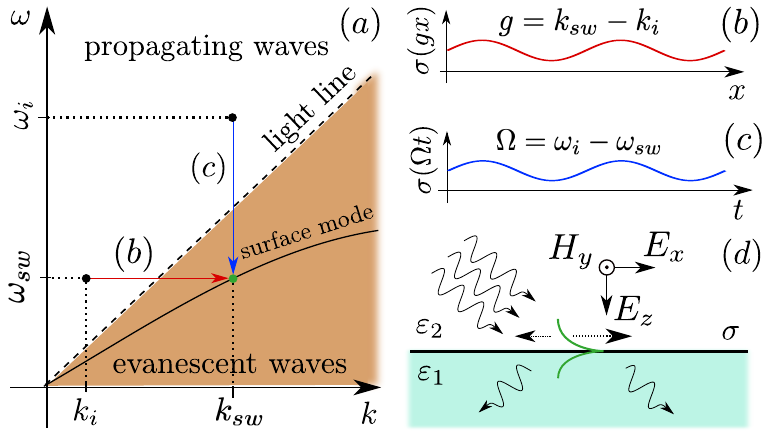}
    \caption{(a,b) Conventional coupling of propagating waves to surface waves is achieved by breaking momentum conservation via a periodic perturbation $\sigma$(x) in space, inducing horizontal transitions of the radiation field,  from the incident wavevector $k_i$ to that of the surface wave $k_{sw}$. Conversely, a periodic perturbation $\sigma(t)$ in time (a,c) breaks energy (frequency) conservation, coupling a wave with incoming frequency $\omega$ to a surface wave with frequency $\omega_{sw}$, thus enabling (d) far-field excitation of surface waves, without the need for in-plane structuring.}
    \label{fig:fig1}
\end{figure}{}

Fig. \ref{fig:fig1} (panels a and b) depicts the conventional scattering process originally observed by Wood: a TM-polarized, propagating wave with in-plane wavevector $k_i$  and frequency $\omega_{sw} = c k_i / \cos(\theta)$, where $\theta$ is the angle measured from the surface plane, impinges on a periodically structured surface which supports a guided mode with wavevector $k_{sw}>\omega_{sw}/c$. If the surface is patterned by a periodic grating whose wavenumber $g = 2\pi/d$, where d is the spatial period, matches the difference between the momentum of the surface wave $k_{sw}$ and that of the photon, $k_i$, then the two modes can be coupled, and a surface wave is launched. This process is described by the red arrow in Fig. 1a.

As a counterpart of this process, if the electromagnetic parameters of a smooth surface are periodically modulated in time with frequency $\Omega$ (Fig. \ref{fig:fig1}a-c), the temporal analogue of a Wood anomaly takes place: an incoming wave with frequency $\omega_i$ and wavevector $k_{sw} = \omega_i\cos(\theta)/c$ is now efficiently coupled to a surface mode if the modulation frequency $\Omega$ matches the difference $\omega_i-\omega_{sw}(k_{sw})$ between the frequency of the incoming photon and that of the surface mode at the incoming wavevector.

Our model system consists of a conductive sheet of infinitesimal thickness, whose dispersive response is described by the Drude equation \cite{maslov2018temporal}:
\begin{align}
 \frac{\mathrm{d} J}{\mathrm{d} t} + \gamma J(x,t) = W_{D}(t) E_x(x,t), \label{eq:drude}
\end{align}{}
for the surface current density $J$, where $W_D(t)$ is the Drude weight, which quantifies the density of charge carriers at any given instant of time, $\gamma$ is a phenomenological dissipation rate, and $E_x$ the in-plane electric field. This model is valid for a general Drude sheet of subwavelength thickness. We assume a harmonic modulation of the Drude weight: $W_D(t) = W_{D,0}(1 + 2\alpha \cos(\Omega t))$, where $\alpha$ and $\Omega$ are the modulation strength and frequency.

Since the system is periodic in time, we can assume Floquet solutions of the form $\psi_{\mathbf{k}}(\mathbf{r},t) = e^{i \mathbf{k}\cdot \mathbf{r}} e^{-i\omega t} \sum \psi_n e^{-in\Omega t}$ for the electric field $\mathbf{E} = E_x \mathbf{\hat{x}} + E_z \mathbf{\hat{z}}$, the magnetic field $\mathbf{H} = H_y \mathbf{\hat{y}}$, and the current $\mathbf{J}= J \mathbf{\hat{x}}$. Solving Maxwell's Equations on the two sides of the current sheet, and imposing continuity of the in-plane electric field $E_x$ and discontinuity of the magnetic field $H_y$ by the surface current $J$, we arrive at the system of equations:
\begin{align}
    \mathbb{D} \mathbf{E}_{x}^{tra} + \mathbb{F} \mathbf{E}_x^{tra} = \mathbf{E}_{x}^{inc}, \label{eq:matrix_equation}
\end{align}
where $\mathbf{E}_x^{inc}$ and $\mathbf{E}_x^{tra}$ are vectors which contain the Fourier amplitudes of the incident and transmitted components of the in-plane electric field, and $\mathbb{D}$ and $\mathbb{F}$ are diagonal and tridiagonal matrices respectively, whose elements read:
\begin{align}
    D_{n,n'} &= \frac{1}{2}[1+ \frac{\varepsilon_1 k_{z,n}^{(2)} }{\varepsilon_2 k_{z,n}^{(1)}}]\delta_{n,n'} \\
    F_{n,n'} &= -\frac{\mu_0 c_0^2 k_{z,n}^{(2)}W_{D,0}[\delta_{n,n'}+\alpha(\delta_{n,n'+1}+\delta_{n,n'-1})]}{2\varepsilon_2(\omega+n\Omega)[\gamma-i(\omega+n\Omega)]} \nonumber \label{eq:bc4},
\end{align}
where $\varepsilon_1$ and $\varepsilon_2$ are the relative permittivities of the lower and upper half spaces [See Fig. \ref{fig:fig1}(d)], $c_0$ is the speed of light in vacuum and $k_{z,n}^{(j)} = \sqrt{\varepsilon_j(\omega+n\Omega)^2/c_0^2-k_x^2}$.

\begin{figure}[t!]
    \centering
    \includegraphics[width = 0.8\columnwidth]{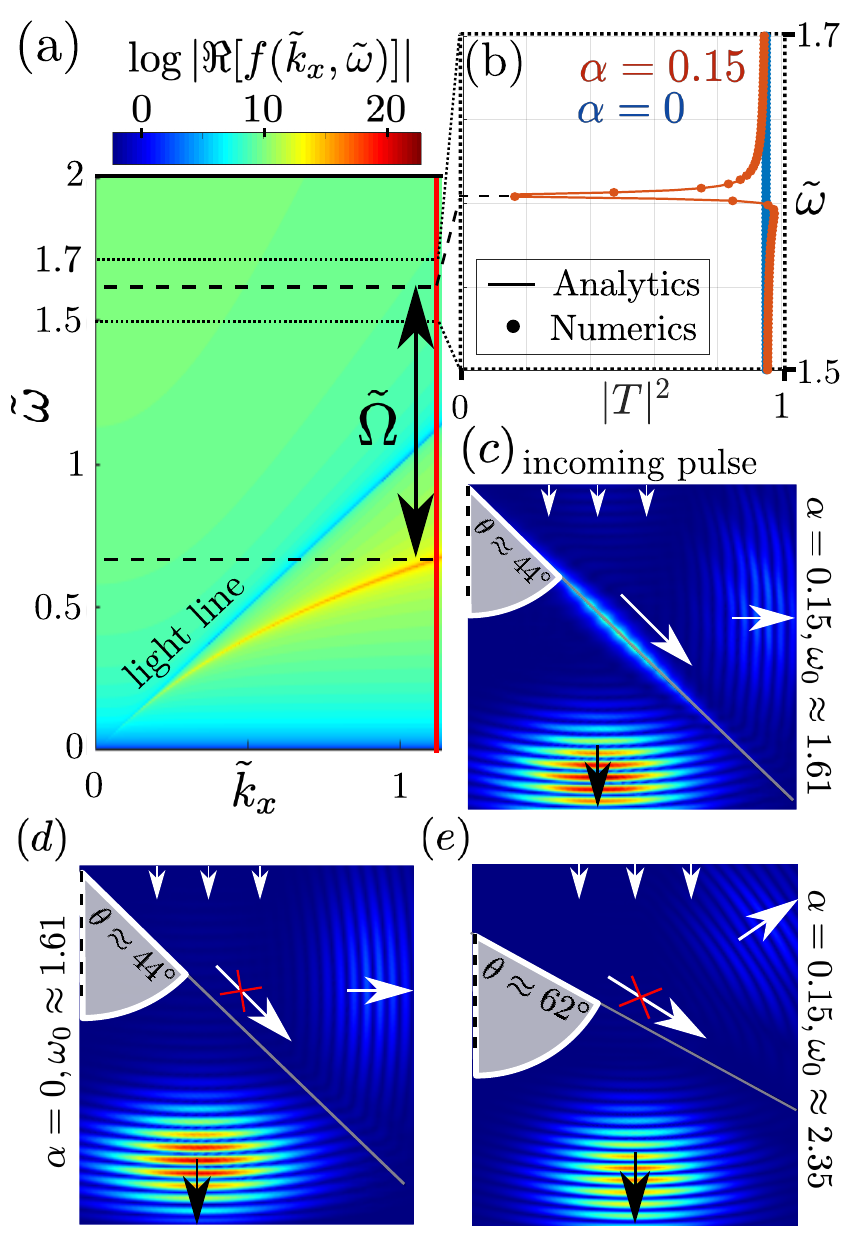}
    \caption{(a) Dispersion relation for surface waves on a current sheet (or deeply subwavelength conducting film) with Drude dispersion (Eq. \ref{eq:disp_rel}). (b) Transmittance through the current-sheet under a time-modulation of its Drude weight at frequency $\tilde\Omega \approx 0.94$, with 30\% modulation amplitude ($\alpha = 0.15$) at fixed in-plane wavevector $\tilde{k}_x \approx 1.12$ (vertical red line); the loss rate $\tilde{\gamma}=0.0011$ [different loss levels considered in Fig. 2 of \cite{supp_mat}]. A surface mode is excited when the frequency of the incoming waves differs from that of the surface wave by the modulation frequency $\tilde\Omega$. Continuous line: analytic theory (3-bands approximation). Dots: exact transfer matrix result. (c-e) FETD simulations of the temporal Wood anomaly predicted in (a-b). (c) A pulse with carrier frequency $\tilde{\omega}_0 \approx 1.61$ impinges on the time-modulated surface ($\alpha=0.15$), successfully exciting a surface wave. In (d), the same pulse impinges on the surface without modulation ($\alpha=0$), and in (e) the modulation is present ($\alpha=0.15$), but the pulse is detuned in frequency, whilst having the same in-plane momentum.}
    \label{fig:fig2}
\end{figure}{}

Defining the matrix on the LHS $\mathbb{T}^{-1}=\mathbb{D}+\mathbb{F}$, we can thus calculate the vector of transmitted amplitudes $\mathbf{E}_x^{tra}=\mathbb{T}\mathbf{E}_x^{inc}$ numerically, by matrix inversion. 

Furthermore, by truncating the matrices in Eq. \ref{eq:matrix_equation} to three modes, where we account for the incoming, propagating wave and the two down-converted modes, and assuming plane-wave incidence and symmetric surrounding media $\varepsilon_1=\varepsilon_2$, we can reduce this problem to the linear system:
\begin{align}
    \begin{pmatrix}
1+i\xi_{-2} & i\alpha\xi_{-2} &  \\
i\alpha\xi_{-1}& 1+i\xi_{-1} & i\alpha\xi_{-1} \\
& i\alpha\xi_{0} & 1+i\xi_{0}
    \end{pmatrix}{}
    \begin{pmatrix}{}
        E_{x,-2}^{tra} \\
        E_{x,-1}^{tra} \\
        E_{x,0}^{tra} 
    \end{pmatrix} = 
    \begin{pmatrix}{}
        0 \\
        0 \\
        E_{x,0}^{inc} 
    \end{pmatrix}
\end{align}{}
which can be solved explicitly to give an approximate analytic solution, accurate in the weak-coupling regime ($\alpha \ll 1$), for the scattering amplitude of the (propagating) transmitted wave:
\begin{align}
    \frac{E_{x,0}^{tra}}{E_{x,0}^{inc}} = \bigg[ \frac{\alpha^2 \xi_0\xi_{-1}}{\bigg(  \frac{\alpha^2 \xi_{-1}\xi_{-2}}{ 1+i\xi_{-2} }+(1+i\xi_{-1}) \bigg) } + (1+i\xi_0) \bigg]^{-1},
\end{align}{} where $\xi_n = \frac{\mu_0 c_0^2k_{z,n}^{(1)}W_{D,0}}{2(\omega+n\Omega)[\gamma-i(\omega+n\Omega)]}$.

Fig. \ref{fig:fig2}(a) shows the dispersion relation for surface modes on a flat, unmodulated Drude sheet of subwavelength thickness \cite{maier2007plasmonics}:
\begin{align}
    f(\tilde{k}_x,\tilde\omega) = \frac{\varepsilon_1}{\sqrt{\tilde{k}_x^2-\varepsilon_1\tilde{\omega}^2}}+ \frac{\varepsilon_2}{\sqrt{\tilde{k}_x^2-\varepsilon_2\tilde{\omega}^2}}
        - \frac{1}{\tilde\omega^2}=0 \label{eq:disp_rel},
\end{align}{}
where we defined dimensionless units $\tilde\omega = \omega/(W_{D,0}Z_0)$ and $\tilde{k}_x= k_x c/(W_{D,0}Z_0)$, where $Z_0=\sqrt{\mu_0/\varepsilon_0}$ is the impedance of free space, and we assumed, for simplicity, that the sheet is surrounded by vacuum. The orange feature shows the dispersion of a surface mode: fixing a wavevector $\tilde{k}_x \approx 1.12$ (red vertical line), we see that a surface mode exists, with frequency $\tilde\omega_{sw}\approx 0.67$; clearly, this mode is inaccessible by any propagating wave, whose dispersion lies within the light cone, unless either in-plane translational symmetry, or, as we set out to show, temporal symmetry, are broken. Assuming a modulation frequency $\tilde{\Omega}\approx 0.94$ and calculating the transmission amplitude for incoming (propagating) waves of different frequencies reveals a clear transmission dip when $\tilde\omega \approx \tilde\omega_{sw}(\tilde{k}_i) + \tilde\Omega \approx 1.61$, showing that incident radiation has coupled to a surface wave. Furthermore, the accuracy of our approximate analytic solution despite up-conversion being neglected highlights how, although the modulation is time-reversal-symmetric, frequency down-conversion strongly dominates the scattering process over high-harmonic generation. This fact highlights the potential of this strategy for the enhancement of parametric down-conversion.

\begin{figure}[t!]
    \centering
    \includegraphics[width = 0.95\columnwidth]{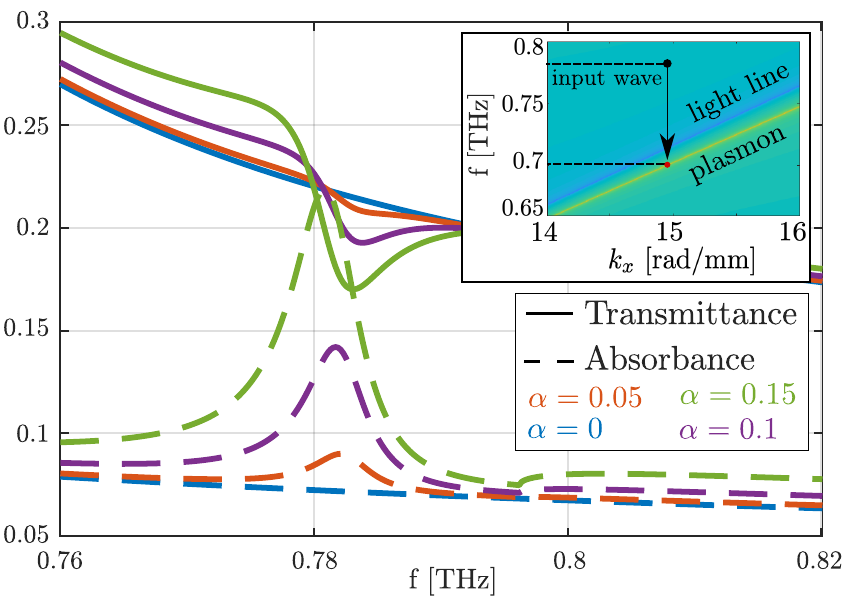}
    \caption{(a) Numerically calculated transmittance (continuous lines) and absorbance (dashed lines) spectra for a graphene sheet whose doping is modulated at a frequency $f_m = \Omega/ 2\pi \times 80$ GHz, with amplitudes of $0$ (blue), 10 (red), 20 (purple) and 30\% (green). The baseline Fermi level $E_F=1$ eV, the electron mobility $m=20'000$ cm$^2$/(V$\cdot$s) and the incident in-plane wavevector $k_x=15$ rad$\cdot$mm$^{-1}$ ($\theta \approx 23^\circ$ at the resonance). The inset shows the transition in the phase plane. }
    \label{fig:fig3}
\end{figure}

In order to verify the frequency-domain results, we then show finite-element-time-domain (FETD) simulations (COMSOL), fully accounting for material dispersion [See Supplemental Material]. We consider an incident pulse with carrier frequency $\omega_0\approx 1.61$, incident at an angle $\theta \approx 44^\circ$, matching in-plane momentum. Fig. \ref{fig:fig3}(c) shows a snapshot of the absolute value of the electric field after the scattering has occurred: The pulse is partly transmitted through and partly reflected from the sheet, leaving behind, however, a surface wave which propagates along the sheet, despite the absence of any spatial inhomogeneity. [Full animations available in the supplementary material]. For completeness, we also show in Fig. \ref{fig:fig3}(d) that no surface wave is excited in the absence of time-modulation under the same illumination, and that (e) the same modulation amplitude $\alpha= 0.15$ does not couple a pulse with detuned carrier frequency $\omega_0 \approx 2.35$ and identical in-plane momentum.

This concept is completely general to any translationally invariant surface which supports a guided mode. Potential photonic implementations are readily possible by using modulated spoof metasurfaces at frequencies in the RF and mm-wave range, where modulation up to speeds of a few GHz are possible \cite{wu2019serrodyne,ramaccia2019phase}. At higher frequencies all-optical approaches may enable this effect to be realised in the terahertz, on graphene, via electro-optic modulation, and potentially in the telecom-band, by exploiting nonlinearities combined with the wide tunability of epsilon-near-zero films such as indium tin oxide \cite{liu2011graphene,li2014ultrafast,woessner2017electrical,alam2016large,shaltout2019spatiotemporal,datta2020low}. Here, we propose a terahertz implementation with surface plasmons in single-layer graphene, which is well within current experimental capabilities \cite{datta2020low}.

Graphene has established itself as a widely tunable material, with reported doping modulation amplitudes of $\approx 38\%$ and $\approx 2.2$ ps response times \cite{li2014ultrafast,tasolamprou2019experimental}, and a number of schemes have been proposed to exploit temporal control of the carrier density of graphene, \cite{wilson2018temporal, ginis2015tunable, galiffi2019broadband}, both by making use of a time-varying bias \cite{correas2018magnetic}. Whilst the quality of graphene has long been known to deteriorate with nanofabrication due to the resulting introduction of defects and edges, surface plasmons in state-of-the-art, homogeneous graphene can achieve measured propagation lengths of tens of microns, surviving up to 50 oscillation periods \cite{ni2018fundamental}.  Could a temporal grating excite graphene plasmons without the need for any surface structure?

Surface plasmons in graphene are generally strongly confined. However, working sufficiently close to the light line offers an opportunity to apply this concept with realistic modulation rates. In Fig. \ref{fig:fig3} we calculate the transmittance and absorbance spectra for incoming THz waves with in-plane wavevector $k_x = 15$ rad/mm incident on a single graphene sheet, whose baseline Fermi level $E_{F,0}=1$ eV, corresponding to a baseline Drude weight $W_{D,0} = \frac{e^2}{\pi \hbar^2}E_{F,0} \approx 120$ GHz/$\Omega$, sinusoidally modulated in time at a frequency $f_m=\Omega/2\pi = 80$ GHz. We assume a conservative electron mobility $m= 20000$ cm$^2$/(V$\cdot$s), corresponding to a loss rate $\gamma = e v_F^2 / m E_F\approx 0.45$ THz, where $v_F=9.5 \times 10^7$ cm/s is the Fermi velocity of the charge carriers. A close-up of the dispersion relation and the coupling process is shown in the inset. From the parameters above, we predict a measurable signal, in transmission, of order ranging between 5\% and 30\%. 

\begin{figure}[t!]
    \centering
    \includegraphics[width = 0.8\columnwidth]{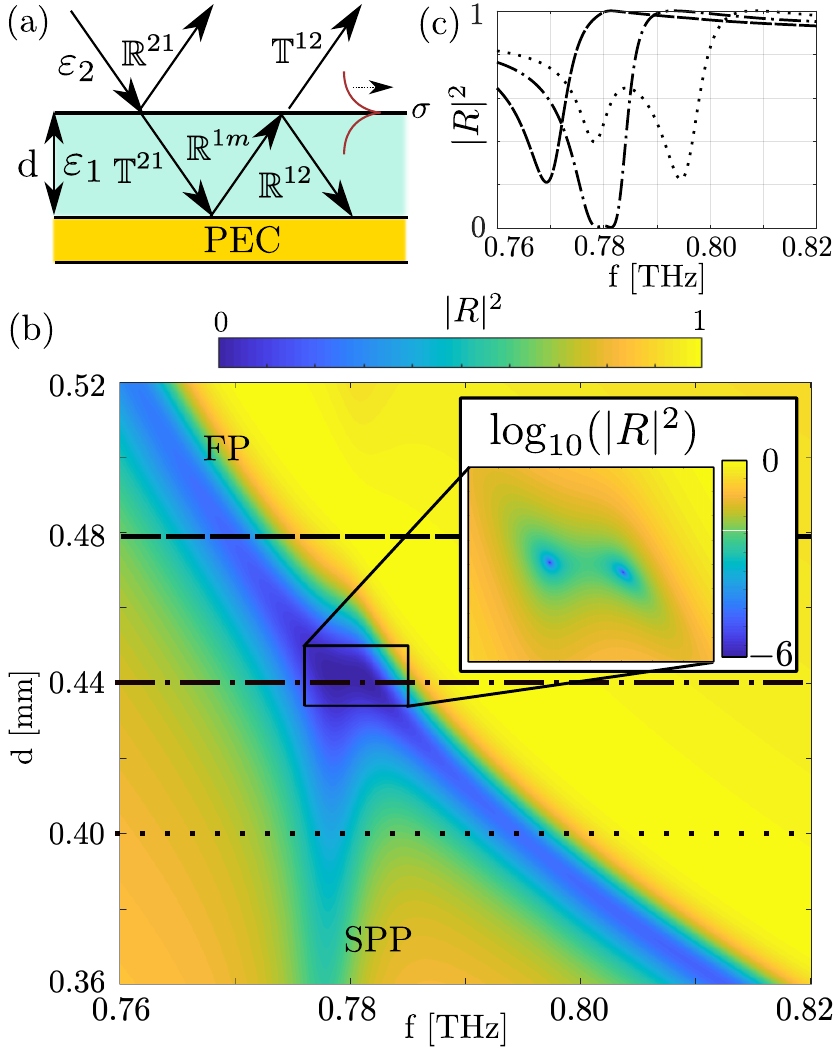}
    \caption{(a) The dynamical excitation mechanism can yield critical coupling to the desired surface mode after including a back-reflector behind the graphene sheet, and coupling the surface plasmon excitation to a Fabry-Perot mode of the cavity, as demonstrated in the surface plot (b), which shows complete suppression of reflectance for an optimal value of frequency and cavity width. The inset shows the anti-crossing between the cavity mode and the surface plasmon, and the three cut-lines at $d=0.40$ (dotted), $0.44$ (dot-dashed), and $0.48$ mm (dashed) are shown in (c).}
    \label{fig:fig4}
\end{figure}{}

The coupling efficiency of this mechanism is measurable, but not perfect. Many graphene setups include a metallic gate, which is typically used to adjust the doping level of the sample \cite{gonccalves2016introduction}. Can we use such a metal-backing to couple the plasmon resonance in Fig. \ref{fig:fig3} to a cavity mode, and thus excite graphene plasmons with unit efficiency? 

Fig. \ref{fig:fig4}(a) illustrates the scattering problem including a back-reflector consisting of perfect electric conductor (PEC) which is a good approximation for gold at terahertz frequencies. The total reflection coefficient is straightforwardly calculated by summing the multiple-scattering series for the amplitudes computed via the previous single-layer transfer-matrix calculation, which gives \cite{markos2008wave}
\begin{align}
    \mathbb{R}^{tot} = \mathbb{R}^{21}+\mathbb{T}^{21}\mathbb{P}(\mathbb{I}-\mathbb{R}^{12}\mathbb{P})^{-1}\mathbb{T}^{12},
\end{align}{} 
where the different matrices containing the reflection and transmission coefficients for the Fourier amplitudes are defined in Fig. \ref{fig:fig4}(a), $\mathbb{I}$ is the identity matrix, and $\mathbb{P}$ is a diagonal matrix with elements $P_{n,n'} = \mathbb{R}^{1m}e^{2ik_{z,n}^{(1)}d}\delta_{n,n'}$, where $\mathbb{R}^{1m}$ is the reflection coefficient at the metal surface, which is simply $-1$ for PEC, while the exponential accounts for the round-trip phase accumulation (or decay) of the $n^{th}$ Fourier mode inside the cavity, which has width $d$. We assume the same parameters for the graphene and the dielectrics as in Fig. \ref{fig:fig3}.

The resulting reflection spectrum is given in Fig. \ref{fig:fig4}(b) as a colour plot against frequency $f$ and cavity width $d$. The oblique reflection dip shows a Fabry-Perot (FP) mode of the cavity, whereas the vertical one shows the surface plasmon mode (SPP). As the two resonances cross, they hybridize, achieving critical coupling between the incoming radiation and the graphene surface plasmon with no breaking of translational symmetry. A blow-up of the anti-crossing arising from the interaction between the two modes is given in the inset in log-scale, demonstrating that critical coupling is achieved. Finally, three cut-lines are shown in (c), corresponding to cavity widths $d=0.40, 0.44$ and $0.48$ mm, showing how the coupling efficiency to the distinct SPP and FP modes (dotted) approaches unity as they strongly interact when $d\approx 0.44$ mm (dot-dashed), and the plasmon becomes dark beyond the anti-crossing (dashed).

In this work we have proposed the temporal counterpart of the Wood anomaly as a general new strategy to couple free-space radiation to surface waves while circumventing the conventional needs for surface fabrication and near-field coupling schemes. Conversely to the momentum coupling introduced by Wood, this scheme relies on the down-conversion of a photon with sufficiently high frequency to match the momentum of a surface mode, whilst being able to propagate in free-space. By exploiting a weak temporal modulation of the impedance of a spatially homogeneous surface, we have shown analytically and numerically how surface waves can be excited on a translationally invariant surface. 

Our model accurately accounts for material dispersion, and we have proposed a terahertz implementation in graphene, which is readily accessible with state-of-the-art experimental setups, showing that critical coupling between radiation and graphene plasmons can be achieved with the addition of a flat back-reflector. To conclude, exploiting the recent introduction of tunable materials, this new coupling scheme may prove a valuable asset in the current struggle to reduce losses in nanophotonics while enabling the highest degree of reconfigurability, as well as ultrafast switching, providing a new perspective in the use of temporal inhomogeneities to enhance light-matter interactions. 

\acknowledgements

The authors thank Dr. Simon Yves, Dr. Riccardo Sapienza and Dr. Stefano Vezzoli for the useful discussions. E.G. was supported through a studentship in the Centre for Doctoral Training on Theory and Simulation of Materials at Imperial College London funded by the EPSRC (EP/L015579/1). P.A.H. acknowledges funding from Funda\c c\~ao para a Ci\^encia e a Tecnologia and Instituto de Telecomunica\c c\~oes under projects CEECIND/03866/2017 and UID/EEA/50008/2020. A.A. acknowledged funds from the Air Force Office of Scientific Research, the Simons Foundation and the National Science Foundation.



\newpage


\newpage


\appendix

\bibliographystyle{unsrt}
\bibliography{DynWoodBiblio}

\begin{thebibliography}{10}

\bibitem{wood1902remarkable}
Robert~Williams Wood.
\newblock On a remarkable case of uneven distribution of light in a diffraction
  grating spectrum.
\newblock {\em Proceedings of the Physical Society of London}, 18(1):269, 1902.

\bibitem{sommerfeld1899ueber}
Arnold Sommerfeld.
\newblock Ueber die fortpflanzung elektrodynamischer wellen l{\"a}ngs eines
  drahtes.
\newblock {\em Annalen der Physik}, 303(2):233--290, 1899.

\bibitem{zenneck1907fortpflanzung}
Jonathan Zenneck.
\newblock {\"U}ber die fortpflanzung ebener elektromagnetischer wellen
  l{\"a}ngs einer ebenen leiterfl{\"a}che und ihre beziehung zur drahtlosen
  telegraphie.
\newblock {\em Annalen der Physik}, 328(10):846--866, 1907.

\bibitem{fano1941theory}
Ugo Fano.
\newblock The theory of anomalous diffraction gratings and of quasi-stationary
  waves on metallic surfaces (sommerfeld’s waves).
\newblock {\em JOSA}, 31(3):213--222, 1941.

\bibitem{ritchie1957plasma}
Rufus~H Ritchie.
\newblock Plasma losses by fast electrons in thin films.
\newblock {\em Physical review}, 106(5):874, 1957.

\bibitem{hessel1965new}
A~Hessel and AA~Oliner.
\newblock A new theory of wood’s anomalies on optical gratings.
\newblock {\em Applied optics}, 4(10):1275--1297, 1965.

\bibitem{raether1988surface}
Heinz Raether.
\newblock Surface plasmons on smooth surfaces.
\newblock In {\em Surface plasmons on smooth and rough surfaces and on
  gratings}, pages 4--39. Springer, 1988.

\bibitem{basov2016polaritons}
DN~Basov, MM~Fogler, and FJ~Garc{\'\i}a De~Abajo.
\newblock Polaritons in van der waals materials.
\newblock {\em Science}, 354(6309):aag1992, 2016.

\bibitem{low2017polaritons}
Tony Low, Andrey Chaves, Joshua~D Caldwell, Anshuman Kumar, Nicholas~X Fang,
  Phaedon Avouris, Tony~F Heinz, Francisco Guinea, Luis Martin-Moreno, and
  Frank Koppens.
\newblock Polaritons in layered two-dimensional materials.
\newblock {\em Nature materials}, 16(2):182--194, 2017.

\bibitem{sievenpiper1999high}
Dan Sievenpiper, Lijun Zhang, Romulo~FJ Broas, Nicholas~G Alexopolous, and Eli
  Yablonovitch.
\newblock High-impedance electromagnetic surfaces with a forbidden frequency
  band.
\newblock {\em IEEE Transactions on Microwave Theory and techniques},
  47(11):2059--2074, 1999.

\bibitem{pendry2004mimicking}
JB~Pendry, L~Martin-Moreno, and FJ~Garcia-Vidal.
\newblock Mimicking surface plasmons with structured surfaces.
\newblock {\em science}, 305(5685):847--848, 2004.

\bibitem{maier2007plasmonics}
Stefan~Alexander Maier.
\newblock {\em Plasmonics: fundamentals and applications}.
\newblock Springer Science \& Business Media, 2007.

\bibitem{renger2009free}
Jan Renger, Romain Quidant, Niek van Hulst, Stefano Palomba, and Lukas Novotny.
\newblock Free-space excitation of propagating surface plasmon polaritons by
  nonlinear four-wave mixing.
\newblock {\em Physical review letters}, 103(26):266802, 2009.

\bibitem{constant2016all}
Thomas~J Constant, Samuel~M Hornett, Darrick~E Chang, and Euan Hendry.
\newblock All-optical generation of surface plasmons in graphene.
\newblock {\em Nature Physics}, 12(2):124--127, 2016.

\bibitem{alam2016large}
M~Zahirul Alam, Israel De~Leon, and Robert~W Boyd.
\newblock Large optical nonlinearity of indium tin oxide in its
  epsilon-near-zero region.
\newblock {\em Science}, 352(6287):795--797, 2016.

\bibitem{shaltout2019spatiotemporal}
Amr~M Shaltout, Vladimir~M Shalaev, and Mark~L Brongersma.
\newblock Spatiotemporal light control with active metasurfaces.
\newblock {\em Science}, 364(6441):eaat3100, 2019.

\bibitem{bruno2020negative}
V~Bruno, C~DeVault, S~Vezzoli, Z~Kudyshev, T~Huq, S~Mignuzzi, A~Jacassi,
  S~Saha, YD~Shah, SA~Maier, et~al.
\newblock Negative refraction in time-varying strongly coupled
  plasmonic-antenna--epsilon-near-zero systems.
\newblock {\em Physical Review Letters}, 124(4):043902, 2020.

\bibitem{datta2020low}
Ipshita Datta, Sang~Hoon Chae, Gaurang~R Bhatt, Mohammad~Amin Tadayon, Baichang
  Li, Yiling Yu, Chibeom Park, Jiwoong Park, Linyou Cao, DN~Basov, et~al.
\newblock Low-loss composite photonic platform based on 2d semiconductor
  monolayers.
\newblock {\em Nature Photonics}, 14(4):256--262, 2020.

\bibitem{sounas2017non}
Dimitrios~L Sounas and Andrea Al{\`u}.
\newblock Non-reciprocal photonics based on time modulation.
\newblock {\em Nature Photonics}, 11(12):774--783, 2017.

\bibitem{ramaccia2019phase}
Davide Ramaccia, Dimitrios~L Sounas, Andrea Al{\`u}, Alessandro Toscano, and
  Filiberto Bilotti.
\newblock Phase-induced frequency conversion and doppler effect with
  time-modulated metasurfaces.
\newblock {\em IEEE Transactions on Antennas and Propagation}, 2019.

\bibitem{caloz2019spacetime}
Christophe Caloz and Zo{\'e}-Lise Deck-L{\'e}ger.
\newblock Spacetime metamaterials, part i: General concepts.
\newblock {\em IEEE Transactions on Antennas and Propagation}, 2019.

\bibitem{cardin2020surface}
Andrew~E Cardin, Sinhara~R Silva, Shai~R Vardeny, Willie~J Padilla, Avadh
  Saxena, Antoinette~J Taylor, Wilton~JM Kort-Kamp, Hou-Tong Chen, Diego~AR
  Dalvit, and Abul~K Azad.
\newblock Surface-wave-assisted nonreciprocity in spatio-temporally modulated
  metasurfaces.
\newblock {\em Nature Communications}, 11(1):1--9, 2020.

\bibitem{pachecopena2020temporal}
Victor Pacheco-Pena and Nader Engheta.
\newblock Temporal antireflection coatings.
\newblock {\em Optica}, 7(4):323--331, 2020.

\bibitem{fleury2016floquet}
Romain Fleury, Alexander~B Khanikaev, and Andrea Alu.
\newblock Floquet topological insulators for sound.
\newblock {\em Nature communications}, 7(1):1--11, 2016.

\bibitem{darabi2020reconfigurable}
Amir Darabi, Xiang Ni, Michael Leamy, and Andrea Alu`.
\newblock Reconfigurable floquet elastodynamic topological insulator based on
  synthetic angular momentum bias.
\newblock {\em Science Advances (in press)}.

\bibitem{trainiti2019time}
Giuseppe Trainiti, Yiwei Xia, Jacopo Marconi, Gabriele Cazzulani, Alper Erturk,
  and Massimo Ruzzene.
\newblock Time-periodic stiffness modulation in elastic metamaterials for
  selective wave filtering: theory and experiment.
\newblock {\em Physical review letters}, 122(12):124301, 2019.

\bibitem{bacot2016time}
Vincent Bacot, Matthieu Labousse, Antonin Eddi, Mathias Fink, and Emmanuel
  Fort.
\newblock Time reversal and holography with spacetime transformations.
\newblock {\em Nature Physics}, 12(10):972--977, 2016.

\bibitem{maslov2018temporal}
AV~Maslov and MI~Bakunov.
\newblock Temporal scattering of a graphene plasmon by a rapid carrier density
  decrease.
\newblock {\em Optica}, 5(12):1508--1515, 2018.

\bibitem{supp_mat}
See supplemental material, which includes refs.~\cite{Wunsch2006}.

\bibitem{wu2019serrodyne}
Zhanni Wu and Anthony Grbic.
\newblock Serrodyne frequency translation using time-modulated metasurfaces.
\newblock {\em IEEE Transactions on Antennas and Propagation}, 2019.

\bibitem{liu2011graphene}
Ming Liu, Xiaobo Yin, Erick Ulin-Avila, Baisong Geng, Thomas Zentgraf, Long Ju,
  Feng Wang, and Xiang Zhang.
\newblock A graphene-based broadband optical modulator.
\newblock {\em Nature}, 474(7349):64--67, 2011.

\bibitem{li2014ultrafast}
Wei Li, Bigeng Chen, Chao Meng, Wei Fang, Yao Xiao, Xiyuan Li, Zhifang Hu,
  Yingxin Xu, Limin Tong, Hongqing Wang, et~al.
\newblock Ultrafast all-optical graphene modulator.
\newblock {\em Nano letters}, 14(2):955--959, 2014.

\bibitem{woessner2017electrical}
Achim Woessner, Yuanda Gao, Iacopo Torre, Mark~B Lundeberg, Cheng Tan, Kenji
  Watanabe, Takashi Taniguchi, Rainer Hillenbrand, James Hone, Marco Polini,
  et~al.
\newblock Electrical 2$\pi$ phase control of infrared light in a 350-nm
  footprint using graphene plasmons.
\newblock {\em Nature Photonics}, 11(7):421--424, 2017.

\bibitem{tasolamprou2019experimental}
Anna~C Tasolamprou, Anastasios~D Koulouklidis, Christina Daskalaki,
  Charalampos~P Mavidis, George Kenanakis, George Deligeorgis, Zacharias
  Viskadourakis, Polina Kuzhir, Stelios Tzortzakis, Maria Kafesaki, et~al.
\newblock Experimental demonstration of ultrafast thz modulation in a
  graphene-based thin film absorber through negative photoinduced conductivity.
\newblock {\em ACS photonics}, 6(3):720--727, 2019.

\bibitem{wilson2018temporal}
Josh Wilson, Fadil Santosa, Misun Min, and Tony Low.
\newblock Temporal control of graphene plasmons.
\newblock {\em Physical Review B}, 98(8):081411, 2018.

\bibitem{ginis2015tunable}
Vincent Ginis, Philippe Tassin, Thomas Koschny, and Costas~M Soukoulis.
\newblock Tunable terahertz frequency comb generation using time-dependent
  graphene sheets.
\newblock {\em Physical Review B}, 91(16):161403, 2015.

\bibitem{galiffi2019broadband}
E~Galiffi, PA~Huidobro, and JB~Pendry.
\newblock Broadband nonreciprocal amplification in luminal metamaterials.
\newblock {\em Physical Review Letters}, 123(20):206101, 2019.

\bibitem{correas2018magnetic}
D~Correas-Serrano, A~Al{\`u}, and JS~Gomez-Diaz.
\newblock Magnetic-free nonreciprocal photonic platform based on time-modulated
  graphene capacitors.
\newblock {\em Physical Review B}, 98(16):165428, 2018.

\bibitem{ni2018fundamental}
GuangXin Ni, AS~McLeod, Zhiyuan Sun, Lei Wang, Lin Xiong, KW~Post, SS~Sunku,
  B-Y Jiang, James Hone, Cory~R Dean, et~al.
\newblock Fundamental limits to graphene plasmonics.
\newblock {\em Nature}, 557(7706):530--533, 2018.

\bibitem{gonccalves2016introduction}
Paulo Andr{\'e}~Dias Gon{\c{c}}alves and Nuno~MR Peres.
\newblock {\em An introduction to graphene plasmonics}.
\newblock World Scientific, 2016.

\bibitem{markos2008wave}
Peter Markos and Costas~M Soukoulis.
\newblock {\em Wave propagation: from electrons to photonic crystals and
  left-handed materials}.
\newblock Princeton University Press, 2008.

\bibitem{Wunsch2006}
B~Wunsch, T.~Stauber, F~Sols, and Francisco Guinea.
\newblock {Dynamical Polarization of Graphene at Finite Doping}.
\newblock {\em New J. Phys.}, 8(12):318--318, dec 2006.

\end{thebibliography}

\end{document}